# OBSERVATION OF REPETITION-RATE DEPENDANT EMISSION FROM AN UN-GATED THERMIONIC CATHODE RF GUN

J. P. Edelen[#], Fermilab, Batavia, IL
Y. Sun, Argonne National Lab, Lemont, IL
J. R. Harris, AFRL, Albuquerque, NM
J. W. Lewellen LANL, Los Alamos, NM

*Abstract*

Recent work at Fermilab in collaboration with the Advanced Photon Source and members of other national labs, designed an experiment to study the relationship between the RF repetition rate and the average current per RF pulse. While existing models anticipate a direct relationship between these two parameters we observed an inverse relationship. We believe this is a result of damage to the barium coating on the cathode surface caused by a change in back-bombardment power that is unaccounted for in the existing theories. These observations shed new light on the challenges and fundamental limitations associated with scaling an un-gated thermionic cathode RF gun to high average current machines.

## INTRODUCTION

Thermionic cathodes are widely known as a robust source of electrons. When used in un-gated RF guns they result in a simple electron injector that is suitable for a wide range of applications for accelerators [1,2,3]. However, one primary drawback of this system is that the cathode will emit whenever the field is appropriately signed. Therefore, some electrons that are emitted late relative to the RF phase will not gain enough energy to exit the cathode cell. These electrons will be decelerated when the field changes sign and eventually accelerated back towards the cathode surface. These so-called back-bombarded electrons deposit their energy on the cathode surface in the form of heat. Because of the cathode's finite thermal mass this additional heating impedes the ability to regulate the output current of the gun. In order to improve our understanding of how changing the back-bombardment heating will affect the beam current, we conducted an experimental study at the Advanced Photon Source Injector Test Stand.

For a gun with a fixed geometry, the back-bombardment power can be varied by changing three parameters: 1) the cathode heater power which changes the cathode current, 2) the peak field in the gap which changes the peak energy of back-bombarded electrons, and 3) the RF duty factor which changes the total amount of energy deposited per second. The cathode heater is slow to respond and therefore changing the heater power does not produce step changes in the back-bombardment power. Changing the peak field in the gap also changes the level of field-enhanced emission causing both a change in the energy of back-bombarded particles and a change in the current, which are difficult to decouple. However, changing the RF repetition rate does not have a slow time response and it does not come with any obvious second order effects.

To study how back-bombardment will impact the output current of the gun, we observed the change in current caused by a change in repetition rate for a variety of conditions. In this paper we will show some initial results for how the output current changed with repetition rate for different values of peak field. We provide an analysis of our measurements and finally a hypothesis for the mechanism that produces the observed dependence of output current on repetition rate.

## EXPIRIMENTAL SETUP

The section of the injector test stand used for this experiment has a 1.5 cell room temperature electron gun, two pairs of quadrupoles, two corrector magnets, a current monitor (Bergoz ICT), a bend, a screen, and a fast-Faraday cup. The gun is powered by a klystron capable of delivering up to 4MW of peak power to the gun. The pulse width of the RF system is 1.05 microseconds and the repetition rate can be varied at uneven intervals between 2Hz and 30Hz (2Hz, 4Hz, 6Hz, 8Hz, 10Hz, 12Hz, 15Hz, 20Hz, 30Hz).

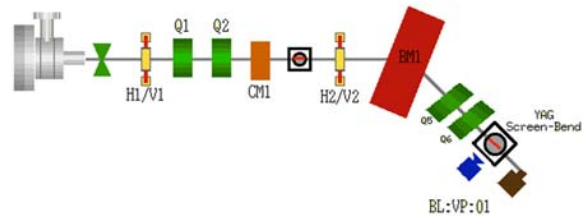

Figure 1: Diagram of the injector test stand used for these experiments, courtesy of the APS control system synoptic display. From left to right, 2.856GHz thermionic gun, gate valve, horizontal/vertical corrector, quadrupole 1, quadrupole 2, current monitor, YAG screen, bending magnet, quadrupole 5, quadrupole 6, YAG screen, and a Faraday cup.

Through the control system we can collect 1Hz data on the average current, magnet settings, forward and reflected power, vacuum pressure, and other diagnostics as needed. Additionally we can periodically collect raw waveform data from the ICT, the faraday cup, and from the RF system using a remotely triggered oscilloscope.

## INITIAL RESULTS

The initial experiment was to vary the repetition rate for several values of peak field at a single heater power.

During follow on testing, we conducted similar measurements at different heater powers. For each test we observed an inverse relationship between repetition rate and output current. Figure 2 shows the output current as a function of time for three different RF power levels. For each of these curves we made a step change in the repetition rate from 30 Hz to 4 Hz at time zero.

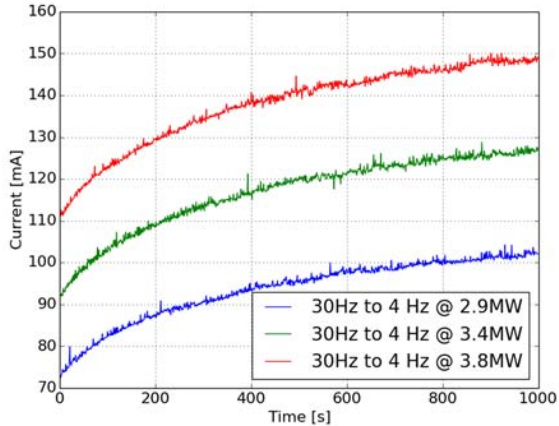

Figure 2: Current as a function of time after decreasing the repetition rate from 30Hz to 4Hz for three different RF power levels

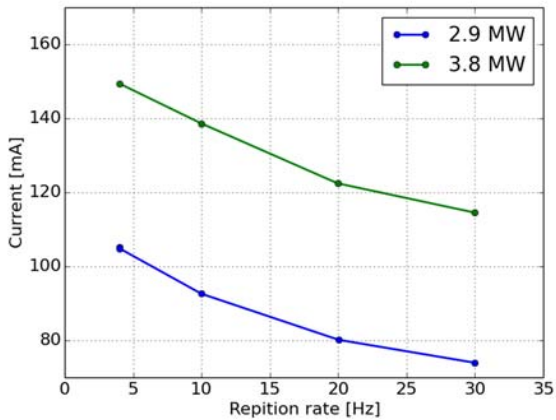

Figure 3: Measured current as a function of repetition rate for two different RF power levels

Figure 3 shows the output current as a function of repetition rate for two RF power levels. Here the output current shown is the average over 30 seconds of data just before switching to the next repetition rate.

The data shown in Figures 2 and 3 are contrary to the existing theories that predict how the output current should vary as a function of repetition rate in the presence of back-bombardment heating [4]. In order to better understand why we observe such a large discrepancy between theory and measurements, we investigate several practical aspects of the experiment that are not accounted for in the theory.

## ANALYSIS OF MEASUREMENTS

A decrease in current as a function of repetition rate could be attributed to a number of effects. First, an increase in repetition rate will cause an increase in the average RF heating in the gun. If the cooling system is too sensitive to changes in RF heating then the cavity will detune causing a decrease in the cavity field for a given forward power, assuming there is no feedback on the RF system. This will subsequently decrease the field enhancement effect and therefore cause a decrease in the current.

We can evaluate this by inspection of the fluctuation of the cavity power. Because this gun does not have a probe signal, the cavity power is computed by taking the forward power minus the reflected. If we are observing a correlated change in the cavity power with the measurements in Figure 2 then it is possible that changes in the field enhancement caused by detuning is the cause of our inverse relationship to repetition rate. Figure 4 shows the fluctuation of the cavity power during the three runs shown in Figure 2.

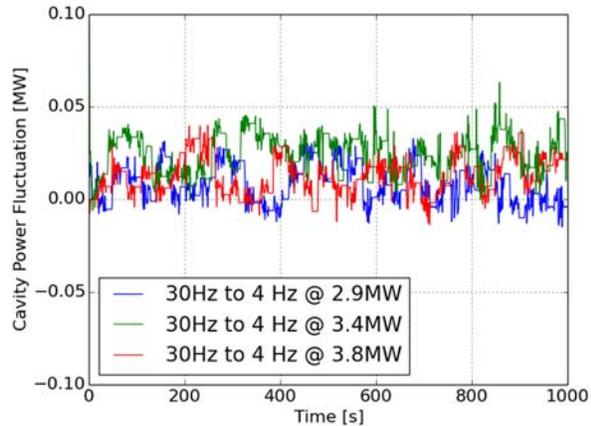

Figure 4: Fluctuation in the work function due to changes in field enhancement during three runs.

Here we can see very small changes in the cavity power and these changes are not well correlated with the measurements in Figure 2. This indicates that we are not seeing a change in the field enhancement.

Another possible cause of the change in beam current is poor control over the cathode heater. We measured the voltage drop and current from the cathode-heater power supply throughout the test to ensure that we were not seeing effects due to changes in heater power. Figure 5 shows the heater power as a function of time during the runs shown in Figure 2.

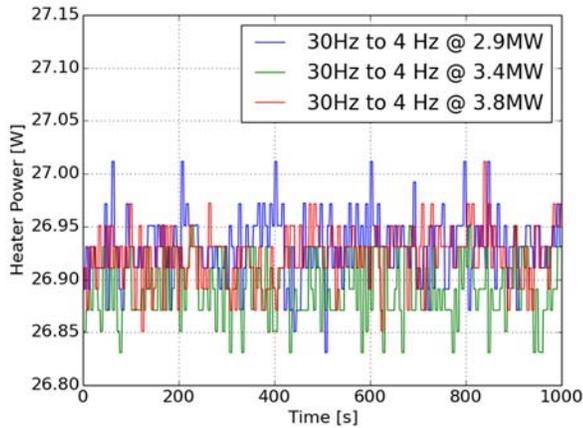

Figure 5: Calculated heater power as a function of time during the three runs shown in Figure 3

While there are small fluctuations in the heater power during the test they are too small to produce the large observed change in output current. Additionally these fluctuations are uncorrelated with the current measurements. It is possible that the experiment is highly sensitive to beam loss that is occurring due to an increase in the current. We did observe some beam loss that could be mitigated through re-tuning of the quads upstream of the ICT, however changes measured at the ICT due to beam-line tuning were on the order of 5%. Additionally we saw the expected correlated increase in the output current due to field enhancement when we increased the RF power from 3 MW to 4 MW. This indicates we have not entered a "negative transconductance"-like regime, where beam interception inside the gun increases faster than the emitted current, leading to a reduction in current leaving the gun despite an increase in current produced at the cathode [6].

Figure 6 shows the beam current as a function of peak field during one of these transitions.

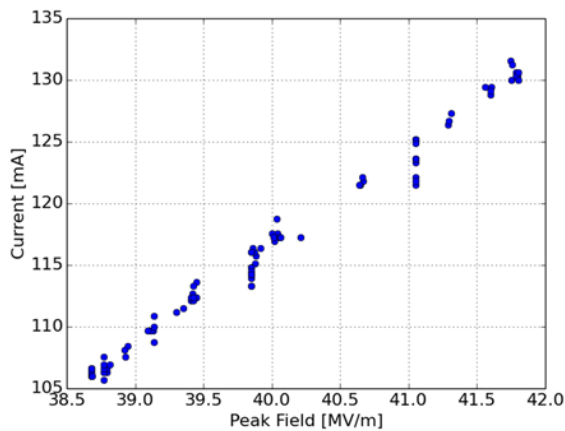

Figure 6: Output current as a function of peak field during a transition from 38 MV/m to 42 MV/m

Finally we propose that the back-bomabardment current is actually damaging the surface of the cathode causing an increase in the work function. In order to study this effect we observed the output current before and after controlled RF trips of different lengths. Figure 7 shows the output current as s function of time during two of these tests.

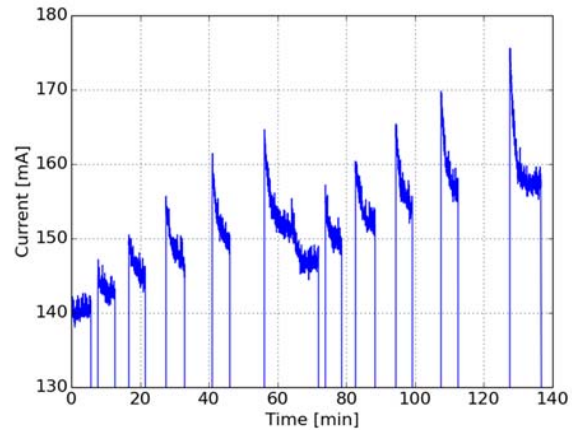

Figure 7: Beam current as a function of time while increasing the RF-off time. At approximately 70 min, we increased the repetition rate to continue the study.

Here we see that as we increase the time the RF is off, the peak current after the RF returns increases. This is consistent with the cathode damage hypothesis, as damage to the surface of our dispenser cathodes will be healed by continued emission of barium from its heated matrix in the period between application of RF; the longer the gun is unpowered, the more fully the cathode should heal, and therefore the larger the increase in current we would expect.

## CONCLUSIONS AND FUTURE WORK

We have observed an unexpected dependence of the output current in a thermionic cathode RF gun on the repetition rate. Existing models predict a strong increase in the output current with repetition rate due to back-bombardment heating while we observed an inverse relationship. Our data indicates that either electron back-bombardment or ion back-bombardment are changing the surface coating of the cathode causing a rise in the work function with repetition rate. This would create the observed inverse relationship of current to repetition rate. Follow on experiments as well as theoretical analysis are currently being conducted in order to better understand our measurements.

## AKNOWLEDGEMENTS